\documentclass[aps,prl,twocolumn,showpacs,
amssymb]{revtex4}
\usepackage{epsfig}
\usepackage{dcolumn}
\usepackage{bm}
\usepackage{amsmath,amssymb}

\begin{document}
\newcommand{\M}{\mathbb{M}}
\newcommand{\R}{\mathbb{R}}
\newcommand{\HI}{\mathbb{H}}
\newcommand{\C}{\mathbb{C}}
\newcommand{\TI}{\mathbb{T}}
\newcommand{\E}{\mathbb{E}}
\def\etal{{\it et al}.} \def\e{{\rm e}} \def\de{\delta}
\def\dd{{\rm d}} \def\ds{\dd s} \def\ep{\epsilon} \def\de{\delta}
\def\goesas{\mathop{\sim}\limits} \def\al{\alpha} \def\vph{\varphi}
\def\Z#1{_{\lower2pt\hbox{$\scriptstyle#1$}}}
\def\X#1{_{\lower2pt\hbox{$\scriptscriptstyle#1$}}}
\def\MM#1{{\cal M}^{#1}} \def\Vm{V(\varphi_0)}
\def\VF{V\Z F} \font\sevenrm=cmr7 \def\ns#1{_{\hbox{\sevenrm #1}}}
\def\Vmin{V\ns{min}} \def\sgn{\,\hbox{sgn}}
\def\coshf{\cosh\left(\chi(u-u\Z0)\right)}
\def\frn#1#2{{\textstyle{#1\over#2}}}
\def\u{{u}}

\title{\bf Accelerating Universes from Compactification on a
Warped Conifold}

\author{Ishwaree P. Neupane}
\affiliation{Department of Physics and Astronomy, University of
Canterbury, Private Bag 4800, Christchurch 8020, New Zealand}
\affiliation{Theory Division, CERN, CH-1211 Geneva 23,
Switzerland\\ E-mail:~{\tt Ishwaree.Neupane@cern.ch} }

\begin{abstract}

We find a cosmological solution corresponding to compactification
of 10D supergravity on a warped conifold that easily circumvents
``no-go" theorem given for a warped or flux compactification,
providing new perspectives for the study of supergravity or
superstring theory in cosmological backgrounds. With fixed volume
moduli of the internal space, the model can explain a physical
Universe undergoing an accelerated expansion in the 4D Einstein
frame, for a sufficiently long time. The solution found in the
limit that the warp factor dependent on the radial coordinate $y$
is extremized (giving a constant warping) is smooth and it
supports a flat four-dimensional Friedmann-Robertson-Walker
cosmology undergoing a period of accelerated expansion with slowly
rolling or stabilized moduli.

\end{abstract}


\maketitle

{\sl Introduction}.-- Recent astronomical data, notably the
observations of high redshift type Ia supernovae~\cite{supernovae}
and measurements of the cosmic microwave background~\cite{WMAP},
not only provide emerging evidence for the ongoing accelerated
expansion of the Universe but also provide support for the concept
of inflation, or a rapid exponential expansion of large magnitude
in a much earlier cosmological epoch. Although it is not difficult
to construct cosmological models that exhibit these features, one
would prefer any such model to be derivable from a fundamental,
and mathematically consistent microscopic theory of (de Sitter)
quantum gravity, such as string theory. Superstring theory lives
in 10 dimensions but we live in a four-dimensional Universe.
Clearly, any attempt to derive a viable cosmology from string or M
theory (compactification) must produce a four-dimensional de
Sitter Universe similar to ours and the size of extra dimensions
should remain much smaller than the physical three dimensions.

The past few years have witnessed significant progress in building
of inflation models within string theory via flux
compactifications of the ten- or eleven-dimensional spacetime of
superstring or M theory with the desire to find models for
late-time cosmology~\cite{KKLT03,TW,IPN03c} supporting a small
positive cosmological constant. If one wishes to stay within the
realm of low energy supergravity models derived from superstrings,
cosmic inflation is ruled out for warped flux compactifications of
classical supergravities on the basis of a ``no-go'' theorem
\cite{nogo,nogo-malda}, which forbids accelerating solutions for
warped (and static) extra dimensions. For a way out, one may
possibly include higher curvature corrections~\cite{Cho-Ish01b} to
the leading order Lagrangian in $\alpha^\prime$ expansion or
extended sources (branes, anti-branes) that are present in string
theory~\cite{GKP01} or even invoke certain non-perturbative
effects (such as, gaugino condensate and Euclidian D3
branes)~\cite{KKLT03}. These all achieve some limited success in
overcoming the no-go theorem. However, there is no good reason to
suppose that all these string effects are available at much lower
energy scale, such as $\rho_{\text vac} \sim {10}^{-3} ~eV$ or
$H\sim 10^{-60}\, M_\text{Pl}$. There is another particular
difficulty in this program in stabilizing the common modulus
associated with the overall shape and size of the internal
Calabi-Yau spaces. Freezing the volume moduli using
non-perturbative dynamics seems beyond anything visible in
supergravity. Time and space are not independent, so any idea that
the geometry of spacetime is fixed and non-dynamical is probably
wrong. The advent of string or M theory in time-dependent
backgrounds is an important and promising subject. It has the
potential to offer a resolution to the dilemma posed by the
observed cosmic acceleration within a natural theoretical
framework. In~\cite{TW} (and generalizations
thereof~\cite{IPN03c}) the no--go theorem was circumvented just by
the choice of negatively curved internal spaces, once the fluxes
are turned off. The restriction on the curvature seems a severe
one, especially given the view that flat space compactifications
on Calabi-Yau {spaces} are among the most natural in string
theory.

An interesting observation in~\cite{IshDavid05a} is that a
time-dependent compactification of classical supergravities with
{\it Ricci-flat} extra spaces, involving certain twists in the
geometry, can give rise to a positive potential in lower
dimensions and hence a period of accelerated expansion in the 4D
Einstein frame. It has been learned that the time-varying volume
moduli with no initial fine tuning among the scalars lead only to
a transient period of cosmic acceleration, except in the case that
we live in a hyperbolic Universe~\cite{IPN03c}. In order to fully
account for the nature of an effective four-dimensional cosmology,
it is important to gain a proper description of spacetime
dependent compactifications (of higher dimensional gravity),
rather than the time-dependent or the space-dependent (warped)
compactification alone. This is because upon dimensional reduction
an internal space of positive curvature gives a negative potential
in time-dependent backgrounds, while it gives a positive potential
in standard warped backgrounds. In view of this observation, all
the studies on flux and time-dependent compactifications (of
string or M theory) to date are either incomplete, or are at best
part of a more complete story.

In this {Letter} we consider a particular model where the internal
spaces have geometries specified by more scalars than just the
volume modulus. The type IIB supergravity theory in ten
dimensions, with a warped 6D conifold geometry provides an example
of this kind, as originally studied by Klebanov and
Tseytlin~\cite{KT00}, and Klebanov and Strassler~\cite{KS00}. In
this model, the internal space is a Ricci-flat 6D cone $Y_6$ whose
base is a 5D Einstein-K\"ahler space, $X_5\equiv T^{1,1}$. The
introduction of branes may be important for constructing gauge
field theories (of the elementary particles) at the tip of a
warped conifold, given a view that both gravitational and
non-gravitational forces can be localized on D3-branes. We show
that the ``no-go'' theorem does not apply to a time-dependent
background even if the extra dimensions are warped.

\medskip
{\sl The model}.-- We shall assume that ten- or eleven-dimensional
supergravity is the relevant starting point. The model below
corresponds to the dimensional reduction to 4 dimensions of type
IIB supergravity, where the spacetime is a warped product of a
six-dimensional space $Y_6$ and $M_4$ ($\equiv {\!\R}^{1,3 }$). In
particular, the 10D metric is
\begin{equation}\label{10Dmetric}
\ds_{10}^2 = h^{-1/2} \tilde{g}_{\mu\nu} {\dd x}^\mu {\dd x}^\nu +
h^{1/2} \dd s_6^2.
\end{equation}
The metric of the large three dimensions (plus time) is
\begin{equation}\label{FRWmetric} \dd x_\mu \dd x^\mu = -
a^{2\de}\dd \u^2 + a^2 (\dd x_1^2+\dd x_2^2+\dd x_3^2),
\end{equation}
where $a\equiv a(\u)$ and $\delta$ is a constant, the choice of
which fixes the nature of the time coordinate $\u$. In the gauge
$\delta=0$, $u$ becomes the proper time $t$. The metric on the
transverse 6D space is
\begin{eqnarray}\label{six-space1}  \dd{s}_6^2 &=&  \e^{2\alpha}
g\, \dd y^2+ \e^{2\beta} k
\sum_{i=1}^{2}\left(d\theta_i^2+\sin^2\theta_i
d\phi_i^2\right)\nonumber \\
&{}& +\, \e^{2\sigma} m (d\psi+f \sum_{i=1}^2 \cos\theta_i
\dd{\phi_i})^2.
\end{eqnarray}
The ranges of the angular coordinates are $0\le \theta_i <\pi$,
$0\le \phi_i < 2\pi$ and $0\le \psi < 4\pi$. We assume that the
moduli parameters other than the volume scalars are stabilized
(frozen); the scalars $\alpha, ~\beta$ and $\sigma $ are functions
of $\u$ (or the proper time $t$), while $g, k, m$ and $h$ are
functions only of the radial coordinate $y$. We will also consider
some examples where $\beta$ and $\sigma$ are functions of the
radial coordinate $y$. As discussed in~\cite{KS00}, due to the
twist along the normal $S^1$ bundle, the model preserves only
$1/4$ of the ${\cal N}=4$ supersymmetries and gives a mass to the
scalar fields. Even though we put in different functions of $y$
for the two 2-spheres and the twisted $S^1$, Einstein's equations
simplify a lot when these functions are proportional, so
henceforth $k(y)\propto m(y)$. The Einstein-frame metric
$g_{E,\mu\nu}$ is related to $\tilde{g}_{\mu\nu}$ via
\begin{equation}
{g}_{E, \mu\nu}=\e^{2\phi} \tilde{g}_{\mu\nu}.
\end{equation}
One must choose the scalars to satisfy ${\alpha}+4 \beta +
{\sigma}\equiv 2\phi$, so that the 4D Newton constant is then
time-independent; $\phi$ is a 4D scalar rather than the 10D
dilaton.

{\sl The background solution}.-- The metric considered by
KT~\cite{KT00} corresponds to the choice $f=1, \alpha=0,
\beta=-\ln\sqrt{6}, \sigma=-\ln 3, g=1$ and $k=y^2$. By turning on
$N$ units of the NS $5$-form flux on $X_5$ and $M$ units of the RR
3-form flux through the $S^3$ of $T^{1,1}$, one finds~\cite{KT00}
\begin{equation}\label{KT-sol}
h(y) = h\Z0+ \frac{L^4}{y^4} \left( 1+ \frac{3 g\Z{s} M^2}{8 \pi
N} \left(1+4 \ln \frac{y}{y_0}\right)\right)
\end{equation}
with $L^4 \equiv 27\pi g\Z{s} N {\alpha^\prime}^{\,2}/4$, which
satisfies the standard quantization conditions:
$(4\pi^2\alpha^\prime)^{-2}\int\Z{T^{1,1}} F_5=N$ and
$(4\pi^2\alpha^\prime)^{-1}\int\Z{S^3} F_3=M$.
The singularity at $y=0$ may be resolved by deforming the
conifold~\cite{KS00,Zayas00a} or by allowing time-dependence to
the internal space.

Let us momentarily set $h\equiv \text{const}$ (or take $y\gg L$),
$g\equiv 1$ and $k\equiv y^2$, which is relevant to finding a pure
time-dependent solution. Einstein's equations admit the following
explicit solution (in the gauge $\delta=3$)
\begin{eqnarray}\label{Ish-sol}
a &=& a_0\, \e^{\pm \, 2 c\Z1 \u}, \qquad
\alpha=c\Z1\u+\alpha_0,\nonumber\\
\beta &=& c\Z1\u+ \alpha_0 - \ln \sqrt{6} = \sigma
+\frac{1}{2}\,\ln \frac{3f^2}{2}.
\end{eqnarray}
For the branch $c_1\u<0$, the size of the internal space shrinks
with time, while the size of the physical three spaces can grow if
we choose the negative exponent. This result is remarkable as it
was impossible for internal spaces with a single (common) volume
modulus.

A few remarks may be relevant before we proceed. We have chosen a
factorizable geometry: the dependence of the warp factor $h$ on
time $t$ (or $u$) has been absorbed into the metric
$\widetilde{g}_{\mu\nu}$ (or the scalars $\alpha, \beta, \sigma$)
as we would like to write the metric in 4D Einstein conformal
frame; a time dependence in $h$ would render it difficult for such
an interpretation. Time-dependent solutions of our sort were
studied in the past, for example, by Kodama and
Uzawa~\cite{Kodama05}. However, it was assumed there rather
implicitly that $\alpha=\beta=\sigma$ and also $g=k=m$. In the
work of Buchel, for example~\cite{Buchel02}, the metric was not
written in the 4D Einstein frame, and also no time-dependence was
allowed for internal spaces. These or other similar assumptions
exhaust some (or all) of the interesting cosmological solutions
that we have found in this Letter.

Kachru {\it et al}.~\cite{KKLT03} proposed to fix the volume
moduli using some non-perturbative dynamics, such as a gaugino
condensate. This is an interesting proposal but such a
construction is model or scheme dependent. For the warped
(conifold) geometry under consideration, the gaugino condensate is
related to the deformation of the conifold, so it is already
visible in the classical geometry, and one does not need
instantons to see the condensate. However, we show that the volume
moduli can be stabilised spontaneously due to a natural expansion
of the Universe, even leading to a transient period of cosmic
acceleration at late times. In general, the volume factors are
dependent on both time and space; in order to write an effective
action in four dimensions, it is necessary to integrate out the
$y$-coordinate. This can be done only if the solutions for both
space- and time-dependent parts of the volume factors are known,
simultaneously.

{\sl Cosmological solution}.-- We shall consider the case
$\alpha=0$ and $g=1$, so as to maintain the interpretation of $y$
as the holographic energy scale. In the zero flux limit of the 10D
Einstein equations, and with the choice $k\equiv y^2$, the
symmetries of the metric ansatz imply that
(i) $4 \beta+ {\sigma}\equiv \varphi(t), \quad h(y)=
{\rho^2}/{y^2}$, (ii) $4{\beta}+{\sigma}=\text{const}\equiv \mu,
\quad h(y)= \sqrt{\lambda + \rho^4/{y^4}}$. Upon dimensional
reduction the first branch above yields
$I=\frac{{\text{vol}(X_5)}}{8\pi G_{10}}\int {d^4 x}\int dy
\sqrt{-g\Z{(4)}}\left( R\Z{(4)} + {\cal L} \right)$, where
$\text{vol}(X_5)$ contains only the space-dependent part,
\begin{eqnarray} {\cal L}& \equiv & K-V= 12
{\dot\beta}^2+\frac{3}{2} {\dot\sigma}^2
+4\dot\beta\dot\sigma -\frac{1}{2}\frac{{h^\prime}^2}{h^3}\,
\e^{-\sigma-4\beta}\nonumber \\
&{}& - \frac{1}{y^2 h} \left(f^2\, \e^{\sigma-8\beta}-4\,
\e^{-\sigma-6\beta} +20\, \e^{-\sigma-4\beta}\right),
\label{KSpoten}
\end{eqnarray}
where $^\prime\equiv d/dy$. The corresponding scalar potential
allows only an anti-de Sitter minimum or it at best describes only
a short period of accelerated expansion due to a relatively large
slope of the potential along the $\beta$-direction. One can easily
modify the form of the potential by introducing a bulk
cosmological term or additional source terms (fluxes, branes) or
even by invoking some particular non-perturbative dynamics, so as
to uplift the AdS minimum and make it a metastable de Sitter
ground state. However, we do not consider this last possibility
here, as it hinders our ability to find analytic solutions.

Instead we consider the second branch, (ii). The type IIB
supergravity equations may be solved by making appropriate ansatz
for the form fields~\cite{KT00,Kodama05}; in the case where the
volume moduli ($\beta, \sigma$) are fixed (or time-independent)
the supergravity equations may be reduced to the
form~\cite{Kodama05}
\begin{eqnarray}\label{sugra-reduced}
R_{\mu\nu} &=& 0, \qquad R_{\mu p}=0, \nonumber\\
\hat{R}_{p q} &\equiv& R_{pq} - \frac{1}{n}  R(X\Z{n}) g_{pq}
(X\Z{n})=0.
\end{eqnarray}
Here $(\mu,\nu)$ run from $1$ to $(10-n)$. In particular, in the
static case, $a\equiv a\Z0$, we define $\beta\equiv \beta(y)$,
$\sigma\equiv \sigma(y)$ (in the metric (\ref{six-space1})) and
take $n=6$. With $k\equiv y^2$, we find
\begin{equation}\label{h-sol-static}
h =h\Z0\, \exp[ c / y^4 ], \qquad \beta= -\sqrt{6(1- 2 c^2/ y^8)}.
\end{equation}
In the large $y$ limit, $h(y\to \infty)\equiv
h_0+L^4/y^4$~\cite{KT00}. All our solutions, both for fixed and
time-dependent volume moduli, satisfy the relation
$\sigma=\beta-\frac{1}{2}\ln (3f^2/2)$, so we write down the
result only for $\beta$. For the solution above only the region
$y^4> \sqrt{2c^2} $ is physical; the singularity at $y=0$ is due
to the choice $k\equiv y^2$, not because of any specific ansatz
for form fields. This is clear also from the deformed conifold
solutions of~\cite{KS00}. To quantify this, suppose that $h\simeq
h\Z0$, without specifying $k(y)$. We find
\begin{equation}
3\left({k^\prime}/{k}\right)^\prime+ k^{-1}\,\e^{-\,2\beta}=0.
\end{equation}
Clearly, if $k\equiv k\Z0 y^2$, then we get $\beta=-\ln\sqrt{6
k\Z0}$, while, if $k\equiv k\Z0\,{\rm sech}(y)$, then
$\beta=-\ln\sqrt{3 k\Z0}+\frac{3}{2} \ln\cosh(y)$, which is
regular everywhere.

The presence of external fluxes would modify the solution for warp
and volume factors as in (\ref{h-sol-static}) for $a\equiv a\Z0$,
or in a more complicated way for $a\equiv a(t)$. Since $R_{yy}=
(h^4 k)^{-1/2}\frac{d}{dy}\left[ h^{9/4} k^3
\frac{d}{dy}\left(h^{-1/4}k^{-5/2}\right)\right]+
\frac{9}{2}\,\frac{d}{dy}(\ln h) \frac{d}{dy}(\ln k)$ and $R_{t
y}=-\frac{5}{2}\dot{\beta} \frac{d}{dy}[\ln (h k)]$, we find
solutions only in the (large volume) limit where $h
k=\text{const}$ (see below), or when the volume moduli are fixed.
Moreover, ${\cal L}\to {\cal L}_{\text{gr}}+{\cal
L}_{\text{flux}}$, where ${\cal L}_{\text{flux}}\propto
\e^{-\,2\phi-2\sigma} (c\Z1^2 \e^{-4\beta}
h^{-2}k^{-3}+\frac{1}{2}\,\e^{-8\beta} h^{-3} k^{-5} K^2-2 c\Z1^2
\e^{2\sigma-4\beta}{F^\prime}^2 h^{-2} k^{-2})$ with $K\equiv
c\Z0+2c\Z1 F(y)$. Under our metric ansatz, equations
$\hat{R}_{pq}=0$ (with $n=5$) are automatically satisfied. To see
the effect of fluxes on the spatial sections of the cosmology, one
can take $k\equiv y^2$ (and hence $h=\sqrt{\lambda+\rho^4/y^4}$).
The explicit solution is
\begin{equation}\label{sol-fixed-moduli}
a(t)=a\Z0\,  \e^{H t}, \quad H^2\equiv \sqrt{\frac{2}{3}} \frac{
|f|\,\rho^8\,\e^{-\,5\beta}}{(\lambda y^4+\rho^4)^{5/2}},
\end{equation}
with an arbitrary (constant) $\beta$. In such a case $R\Z{pq}
(X\Z5)$ is supported by five-form (and self-dual three-form)
fluxes on $T^{1,1}$. The above solution is stable as long as the
volume moduli are fixed. For slowly rolling moduli, any such de
Sitter phase would be only metastable. This example demonstrates
that it is indeed possible to maintain fixed volume modulus of the
internal space while the spatial sections of the cosmology undergo
a de Sitter expansion.

{\sl A constant warping}.-- In the large volume limit, when the
backreaction of the fluxes on Einstein's equations can be ignored
(since their contribution to the stress tensor is volume
suppressed), the warp factor is minimized. This particular case
may be related to the large $\tau$ limit of the resolved conifold
metrics in~\cite{Zayas00a}, for which $h\to h(\tau)\equiv h_0+
h_1\,e^{-4\tau/3}$; $k(\tau)$ and $m(\tau)$ also take their
extremized values. The field equations reduce to equations of
motion, and a constraint, for the variables ($\alpha, \beta,
\sigma$) that are the Euler-Lagrange equations of the effective
Lagrangian ${\cal L}=
\frac{3}{2}\,\dot\alpha^2+12\dot\beta^2+\frac{3}{2}\,\dot\sigma^2
+4\dot\alpha\dot\beta+4\dot\beta\dot\sigma+\frac{1}{2}\,\dot\alpha\dot\sigma
-\frac{\e^{-\,2\phi-4\beta}}{h_0} \left( f^2\,\e^{2\sigma}-4
\e^{2\beta}\right)$. The explicit solution is
\begin{eqnarray}\label{sol-rolling-moduli}
a&=& \e^{\,\zeta \u}\, (\cosh\chi\u)^{-5/8},\quad
\alpha= c\Z1 \u + c_0,\nonumber \\
\sigma &=&  -\frac{1}{4} \ln \cosh \chi \u + c\Z2 \u +
c\Z3=\beta-\frac{1}{2}\ln \frac{3f^2}{2}
\end{eqnarray}
(in the gauge $\delta=3$), where $\zeta \equiv (c\Z1+7 c\Z2)/6$
and
\begin{eqnarray}
h_0= \frac{64}{81}\,\frac{ \e^{-c\Z0-7c\Z3}}{f^6\chi^2},~ \chi^2
\equiv \frac{16}{15}\left(c\Z1^2+2 c\Z1 c\Z2 +7 c\Z2^2\right).
\end{eqnarray}
The four-dimensional cosmic time $t$ is defined by $\dd t=\pm\,
a^3\,\dd u$. It follows that this solution exhibits a period of
accelerated expansion ($\dot{a}>0, ~\ddot{a}>0$) in the 4D
Einstein frame, provided that $2-3\sqrt{2}< c\Z1/c\Z2
<2+3\sqrt{2}$. From a purely metric point of view, all the
solutions with $|f| > 0$ are non-singular. The constants $c\Z0$
and $c\Z3$ may be set to zero using a shift-symmetry in $u$, or
alternatively, can be absorbed into $g$ and $k$ so that each
becomes unity even if they are assigned different values
initially. The scalars $\beta$ and $\sigma$ can be stabilized by
requiring that
$$
c\Z1/c\Z2=2 \quad \text{or} \quad  c\Z1/c\Z2 =- 4. $$ The scale
factor then evolves as $a\sim \left(\e^{-\,c\Z2 u}+\e^{4 c\Z2
u}\right)$ or as $a \sim \left(\e^{-\, 2 c\Z2 u}+\e^{3 c\Z2
\u}\right)$. In the first case the Universe still experiences a
short period of accelerated acceleration before the volume scalars
$\beta$ and $\sigma$ attain nearly fixed values, in which limit
$V\propto \e^{-\,\alpha}$. One takes $c\Z2 u>0$, so that the
physical three spaces expand faster, in the conventional manner.
The radial modulus associated with $\!\R^1$ expands in the first
case, providing a ``4+1+compact space" type background, while it
shrinks in the second case, providing a ``3+1+compact space" type
background. Thus the expansion of $3+1$ spacetime and contraction
(or slow expansion) of the six extra dimensions can fundamentally
be a natural phenomenon. Similar results exist in other versions
of supergravity or string theory.

Consider ten-dimensional {\sl type IIA} supergravity which already
has a (positive) cosmological term~\cite{nogo-malda}:
$I_{gr}=\frac{1}{8\pi G_{10}} \int \sqrt{-g_{10}}\, (R-2\Lambda)$.
In the case $h(y) \to h\Z0$, the 10D Einstein equations are solved
by
\begin{eqnarray}
&&k=\frac{1}{4\nu^2}\,\left(k\Z1 \sin(\nu\, y) -k\Z2 \cos (\nu\,
y)\right)^2,~
\nu \equiv \sqrt{{\sqrt{h\Z0}\Lambda}/{20}}, \nonumber\\
&&a=a\Z0\,\e^{H t},\quad H \equiv \e^{-2\beta-\sigma/2}\,
\sqrt{{\Lambda}/{(12\sqrt{h\Z0})}},
\nonumber \\
&&\beta = \frac{1}{2} \ln
\frac{2}{3(k\Z1^2+k\Z2^2)}=\sigma+\frac{1}{2}\ln \frac{3f^2}{2},
\end{eqnarray}
where $k\Z1, k\Z2$ are arbitrary constants. This solution clearly
supports an accelerated expansion for $\Lambda>0$. It would be
interesting to see a generalization of this result in the case
$h\equiv h(y)$.

We conclude the Letter with a short summary of the results. It has
been a difficult problem to construct accelerating cosmologies
from toroidal or spherical compactification of string or
supergravity theory with stabilized or slowly rolling volume
moduli. In this Letter we have shown that it is possible to
construct an effective four-dimensional cosmology undergoing one
or more periods of accelerated expansion in the general setting of
10D supergravity compactified on a warped 6D conifold, with or
without external fluxes. Allowing time dependence in the warped
conifold solutions is an excellent route for studying aspects of
de Sitter Universe via string compactifications. We considered
explicit cosmologies that arise in models of gravity which
correspond to the dimensional reduction to 4 dimensions of 10-d
supergravity. It is remarkable that a model with so many
attractive features can arise from a simple compactification of
type IIB (as well as type IIA) supergravity on a warped conifold.
Further generalizations of the solutions discovered in this Letter
are also possible, including the case where $g=g(y)$. Finally we
note that for slowly rolling moduli, the effective potential,
$V(\phi)$, can vary slowly with time, while for fixed volume
moduli, it acts purely as a cosmological term (cf.
(\ref{sol-fixed-moduli})); thus the model could satisfy the solar
system test and other constraints from cosmology.

I wish to thank Igor Klebanov for constructive remarks and
recommendations on the draft, and Chiang-Mei Chen, Chris Herzog,
Pei-Ming Ho, David Wiltshire for helpful discussions. This
research is supported by the Foundation for Research, Science and
Technology (NZ) under Research Grant No. E5229.


\end{document}